\documentclass[12pt]{iopart}
\usepackage{hyperref}
\usepackage{iopams}
\usepackage{graphicx}

\newcommand{\unit}[1]{\,{\rm #1}}

\providecommand{\boldsymbol}[1]{\mbox{\boldmath $#1$}}

\providecommand{\tabularnewline}{\\}

\begin{document}

\title[Compact Toroid Fueling]{Ideal Magnetohydrodynamic Simulations of Low Beta Compact Toroid Injection into a
Hot Strongly Magnetized Plasma}
\author{Wei Liu$^1$, Scott C. Hsu$^2$, Hui Li$^1$}
\address{$^1$ Theoretical Division, Los Alamos National Laboratory, Los Alamos, NM, USA, 87545}

\address{$^2$ Physics Division, Los Alamos National Laboratory, Los Alamos, NM, USA, 87545}
\ead{wliu@lanl.gov}

\begin{abstract}
We present results from three-dimensional ideal magnetohydrodynamic
simulations of low $\beta$ compact toroid (CT) injection into a hot strongly
magnetized plasma, with the aim of providing insight
into CT fueling of a tokamak with parameters relevant for ITER (International
Thermonuclear Experimental Reactor).  A regime is identified in terms of
CT injection speed and CT-to-background magnetic field ratio that
appears promising for precise core fueling.  
Shock-dominated regimes, which are probably unfavorable for
tokamak fueling, are also identified.  The CT penetration depth is proportional to the CT
injection speed and density.  The entire CT evolution can be
divided into three stages:  (1)~initial penetration, (2)~compression in the
direction of propagation, and reconnection with the background magnetic field, and (3)~coming to rest and spreading in the
direction perpendicular to injection.  Tilting of the CT is not
observed due to the fast transit time of the CT across the background plasma.

\end{abstract}

\submitto{Nuclear Fusion}

\maketitle

\section{Introduction}

It is important to deliver fuel into the core of a tokamak fusion
plasma to maintain steady-state operation, achieve more efficient
utilization of deuterium-tritium fuel, and optimize the energy
confinement time \cite{pp88}. Several fueling schemes have been
proposed, such as edge gas puffing, pellet
injection \cite{mfes79}, and compact toroid (CT)
injection \cite{phh88}.  Among them, CT fueling is considered to be
the most promising method for core fueling because the injection speed
via this method is far higher than those of the other methods.
Although extensive worldwide efforts have been devoted to study CT
fueling theoretically \cite{pp88,phh88,nw91},
numerically \cite{shk00,swsh00,shk01,shk01a}, and
experimentally \cite{bb90,bb92,rmqs94,rmhsa97,rmpsm97,ghbcff97,ffka97,sh97,ddlpgc98,
rg98,hrm99,ofn99,hmbe00,sh99,xhs04,noyf05,lxsh06,om08,rr08}, the dynamics of core
CT fueling of large devices like ITER (International Thermonuclear
Experimental Reactor) \cite{iaea01} is not well understood.  CT
injection has the potential to deposit fuel in a {\it controlled}
manner at any point in the machine, from the edge to the core.
Tangential (toroidal) injection can impart momentum for improving
plasma $\beta$ and stability \cite{rr06}. In a burning plasma
device with only radio-frequency (rf) for auxiliary current drive, a
CT injection system may be the only internal profile control tool for
optimizing bootstrap current and maintaining optimized fusion burn
conditions. CT fueling also provides a good chance to study core
transport in present machines, helium ash removal, and Edge
Localized Mode (ELM) \cite{wf82} control.

In this work, we employ a simple idealized model of a low $\beta$ CT
propagating into a uniform slab plasma with a uniform magnetic field
perpendicular to the CT injection direction, mimicking CT fueling into a tokamak with infinite aspect ratio.  This model helps us
identify different regimes of operation in terms of CT injection
speed, density, and magnetic field strength, as well as understand the
essential physics occurring during CT injection.  More realistic
scenarios, including the use of realistic tokamak profiles and
geometry in the background plasma, as well as high $\beta$ CT's and dense plasma jets, are
planned for follow-on research.  Compared to past work on CT injection
simulations, we have investigated new regimes especially in terms of
higher injection velocity and a more ITER-relevant ratio (at least for
low $\beta$ CT's such as spheromaks) of CT-to-background magnetic
field ($\sim 0.1$).  Simulations with higher injection velocity were
made possible by the shock-handling capability of our
three-dimensional (3D) ideal magnetohydrodynamic (MHD)
code \cite{llc08}. The lower CT-to-background magnetic field
ratios, compared to past work, was enabled by the higher resolution of
our code which allowed the boundary layer between the CT and
background plasma to be properly resolved. 

The paper is organized as follows.  In Sec.~\ref{setup}, we describe
the the problem setup including initialization of the CT and background
slab plasma and the numerical model. We present the simulation results in Sec.~\ref{result},
and our conclusions and implications for future CT fueling experiments
are given in Sec.~\ref{conclusion}.

\section{Problem setup and numerical model}\label{setup}
A low $\beta$ CT with spherical radius $r_b=1$, centered initially at
$x_b=0$, $y_b=0$ and $z_b=z_{b,0}=-12$, is injected along the $z$ axis
into a lower density background plasma with injection velocity $v_{\rm
inj}$ (see Figure~\ref{config}).  The basic model assumptions and
numerical treatments are briefly summarized here; they are essentially
the same as those in Li et al.\cite{llf06} where more details are given. This code uses high-order Godunov-type finite-volume numerical methods. These methods conservatively update the zone-averaged fluid and magnetic field quantities based on estimated advective fluxes of mass, momentum, energy, and magnetic field at the zone interface \cite{llf06}. The divergence-free condition of the magnetic field is ensured by a constrained transport (CT) scheme \cite{bs99}.  All simulations were performed on the parallel Linux
clusters at Los Alamos National Laboratory. The nonlinear
system of time-dependent ideal MHD equations in 3D Cartesian
coordinates $(x,y,z)$ is given here:
\begin{eqnarray}
\frac{\partial \rho}{\partial t}+\nabla\cdot(\rho\vec{v})&=&0\,,\\
\frac{\partial(\rho\vec{v})}{\partial
t}+\nabla\cdot\left(\rho\vec{v}\vec{v}+(p+\frac{B^2}{2})\mathbf{I}-\vec{B}\vec{B}\right)&=&0\,,\\
\frac{\partial E}{\partial
t}+\nabla\cdot\left[\left(E+p+\frac{B^2}{2}\right)\vec{v}-\vec{B}(\vec{v}\cdot\vec{B})\right]&=&0\,,\\
\frac{\partial \vec{B}}{\partial
t}-\nabla\times(\vec{v}\times\vec{B})&=&0\,,
\end{eqnarray}
in which $\rho$, $p$, $\vec{v}$, $\vec{B}$ and E are the density,
(gas) pressure, flow velocity, magnetic field, and total energy,
respectively. $\mathbf{I}$ is the unit diagonal tensor. The total
energy is $E=p/(\gamma-1)+\rho v^2/2+B^2/2$, where $\gamma$ is the
ratio of the specific heats. Note that a factor of $\sqrt{4\pi}$ has
been absorbed into the scaling for both the magnetic field $\vec{B}$
and current density $\vec{j}$.   It should be noted that
the details of effects such as reconnection and heat evolution could not
be addressed accurately due to the ideal MHD model and the use of a
simplified energy equation.

It is well established empirically in coaxial gun spheromak
experiments that, under proper conditions, a spheromak ``magnetic
bubble" (a low $\beta$ CT) will be formed by the gun
discharge \cite{bp00}. In our simulations, we do not model the CT
formation process and instead start with a pre-formed CT moving toward
the background slab plasma at speed $v_{\rm inj}$.  The stationary
background plasma is composed of a slab plasma confined by a uniform
background magnetic field $B_p$ in the $x$ direction.  For simplicity,
we assume the background plasma has uniform initial number density
$\rho_{p}=0.1$ and uniform initial temperature $T_{p}=0.1$.

In the simulations reported here, the CT structure is similar to the
one given in Liu et al.\cite{lhlll08}. The density profile of the CT plasma with
radius $r_b=1$ is given by
\[
\rho_b\propto r_c^2\exp[-r_c^2-(z_c-z_b)^2],
\]
up to a normalization coefficient and a uniform temperature
$T_{b}$, where $r_c=\sqrt{x^2+y^2}$ and $z_c=z$ (see
Figure~\ref{config}).  The density profile used here has its peak
shifted from the center of the CT, approximating a spheromak.

The CT magnetic field is determined by three key quantities: the
length scale of the bubble magnetic field $r_B=2$, the amount of
poloidal flux $\Psi_p$, and the index $\alpha$, which is the ratio of
the CT toroidal to poloidal magnetic fields. For simplicity, the CT
magnetic field $\vec{B}_{b}$ is also assumed to be axisymmetric.
The poloidal flux function $\Psi_p$ is specified as
\begin{equation}
\Psi_p\propto r_c^2\exp[-r_c^2-(z_c-z_b)^2]\,.
\end{equation}
The poloidal fields, up to a normalization coefficient, are
\begin{equation}
B_{{b},r_c}=-\frac{1}{r_c}\frac{\partial \Psi_p}{\partial
z_c},\quad B_{{b},z_c}= \frac{1}{r_c}\frac{\partial
\Psi_p}{\partial r_c},
\end{equation}
while the toroidal magnetic field is
\begin{equation}
B_{{b},\varphi_c}=\frac{\alpha \Psi_p}{r_c}=\alpha
r_c\exp[-r_c^2-(z_c-z_b)^2]\,.
\end{equation}
The azimuthal component of the CT Lorentz force is zero, but the total
azimuthal Lorentz force due to the \emph{combined} fields and currents
of the CT may be non-zero.

The CT also has uniform injection velocity $v_{\rm inj}$ and uniform
rotation angular speed $\omega$. In this paper the ratio of the CT's
toroidal to poloidal magnetic fields $\alpha$, the rotation speed of
the CT $\omega$, and the specific heat $\gamma$ are taken to be
$\sqrt{10}$, $0$ and $5/3$, respectively.  Physical quantities are
normalized by the characteristic system length scale
$R_0=10\;\unit{cm}$, density
$\rho_0=7.77\times10^{-9}\;\unit{g/cm^{-3}}$ (corresponding to plasma
number density, which is the sum of electron and ion number density,
of $18.6\times10^{14}\;\unit{cm^{-3}}$), and velocity
$V_{0}=1.7\times10^{8}\;\unit{cm\;s^{-1}}$. Other quantities are
normalized as: time $t=1$ gives $R_0/V_0=5.9\times10^{-8}\;\unit{s}$,
magnetic field $B=1$ gives
$(4\pi\rho_0V_{0}^2)^{1/2}=5.3\times10^{4}\;\unit{G}$, and energy
$E=1$ gives $\rho_0 V_{0}^2R_0^3=2.24\times10^{11}\;\unit{ergs}$.

The boundary conditions are all perfectly conducting in the $y$ and
$z$ directions except at the port where the CT is injected, while in
the $x$ direction outflow boundary conditions are employed in order to
mimic the toroidal geometry of a tokamak. Since the toroidal dimension is much larger than the poloidal dimension in a real tokamak and we focus on the early stage of the CT evolution before one toroidal propagation time of the Alfv\'en wave induced by the CT propagation, the out-flowing boundary condition is more appropriate than the periodic boundary condition used in Suzuki et al. \cite{shk01}. In order to minimize the influence of the entrance port, the port will be switched on when the top of the CT reaches the bottom boundary at $t=2/v_{\rm inj}$ and it will be switched off at $t=5/v_{\rm inj}$ after the CT has fully entered the computation domain ($t\sim4/v_{\rm inj}$). Suzuki et al.\ have pointed
out that the boundary condition in the background magnetic field ($x$)
direction is important, \emph{i.e.}, magnetic reconnection has more
influence on CT deceleration with perfectly conducting boundary
conditions than with stress-free boundary conditions such as outflow
and periodic boundary conditions \cite{shk01}. The total computational
domain is $|x|\le9$, $|y|\le9$, and $|z|\le9$, corresponding to a
$(180\;\unit{cm})^3$ box in actual length units (assuming the physical
dimension of the injected CT radius $r_b=10\;\unit{cm}$). The
numerical resolution used here is $400\times400\times800$, where the
grid points are assigned uniformly in the $x$, $y$, and $z$
directions. A cell $\delta x$ $(=\delta y=2\delta z=0.045$)
corresponds to $0.45\;\unit{cm}$.  Since the plasma skin depth and ion gyroradius based on the parameters of ITER are no more than $\delta x$, the simulations based on an MHD model are appropriate. 

\section{Results}\label{result}

In this section we present ideal MHD simulation results on the
injection of a low $\beta$ CT into a hot strongly magnetized plasma.  We organize our
results into three topics:  (1)~parameter regimes, in terms of $v_{\rm
inj}$ and ratio of CT-to-background magnetic field, of CT evolution
including the identification of a promising regime for ITER-relevant
precise core fueling, (2)~detailed description of the CT evolution for
the ITER-relevant regime,  and (3)~dependence of the CT penetration
depth on $v_{\rm inj}$ and the initial CT density.


\subsection{Parameter regimes of CT injection}

Based on our simulation results, we find that the evolution of the
injected CT depends predominantly on the initial injection speed
$v_{\rm inj}$ and the initial ratio of CT-to-background
magnetic field.  As shown in Figure~\ref{scheme}, there are
several qualitatively distinct regimes of CT injection in terms of the
above two parameters.  

First, below a threshold injection speed $V_{\rm L}$, the CT is unable
to penetrate the background plasma at all (see \emph{left} panel of Figure~\ref{nonlocal}). The conducting sphere
(CS) model \cite{pp88} requires that
the initial CT kinematic energy exceeds the background
magnetic field energy excluded by the CT volume,
\begin{equation}
\label{cs}
\frac{1}{2}\rho_b v_{\rm inj}^2>\frac{1}{2}B_p^2\rightarrow v_{\rm inj}>V_{\rm AC}=
\frac{B_p}{\sqrt{\rho_b}}\,.
\end{equation}
Our results show that $V_{\rm L}$, which is inferred ($\sim0.5$) from extrapolation of the data shown in Fig.~\ref{penetration}(a), is less than $V_{\rm AC}$. This is because of the compression of the CT during the penetration, which is ignored by the CS model (see discussion of \S\ref{overview}), \emph{i.e.}, the CT actually excludes less volume than its initial volume, therefore leading to a smaller initial injection speed threshold. And since this compression is related to the background plasma magnetic field (see discussion in \S\ref{overview}), this injection speed threshold is thus dependent on the background plasma magnetic field. However, $V_{\rm AC}$ derived from the CS model still gives a rough estimate of the lower limit of the injection speed needed for penetration. 

Second, above a different threshold injection speed $V_{\rm AP}$, a
strong shock and wavefront are observed to develop ahead of the CT,
dominating the system evolution (see \emph{middle} panel of Figure~\ref{nonlocal}).  This threshold is
determined by the condition when $v_{\rm inj}$ exceeds the Alfv\'en
speed of the background plasma, {\em i.e.}, $v_{\rm inj}>V_{\rm
AP}=B_{p}/\sqrt{\rho_p}$.  This shock/wavefront-dominated regime is
probably not favorable for CT fueling because it may strongly perturb
the tokamak equilibrium.  Also, in this case the CT deposition is
highly nonlocal (\emph{middle} panel of Figure~\ref{nonlocal}). For
comparison, the result ($v_{\rm inj}=1.1$) without shock/wavefront
development is also presented (\emph{right} panel of
Figure~\ref{nonlocal}) ($V_{\rm AP}=3.16$ in this case).

Third, in the regime $V_{\rm L}<v_{\rm inj}\lesssim V_{\rm AP}$,  CT
evolution is further determined by the ratio $S_{r}=B_b/B_p$ of the CT
field $B_b$ to the background plasma magnetic field $B_p$.  If $S_r
\gg 10$, a large non-zero initial Lorentz force results in strong CT
expansion that also leads to the development of a shock and a wavefront
that dominate the system evolution \cite{lhlll08}. As stated above,
this does not favor controlled plasma fueling.  In the regime $1 < S_r
\lesssim 10$, which is the regime Suzuki et al.\ have discussed
extensively \cite{shk00,swsh00,shk01}, the CT is decelerated by both
the magnetic pressure and magnetic tension forces.  In this regime, the CT
tilts while reconnection occurs between the CT and background plasma
magnetic fields. Suzuki et al.\ proposed the Non-slipping Conducting
Sphere (NS) model \cite{shk00,swsh00,shk01}, which matches their
simulation results of vertical injection without a magnetic field gradient pretty well \cite{shk01}.  Our simulation results in this regime
verify their conclusions.  Because it is difficult for the CT field
$B_b$ generated by a co-axial gun to be larger than the tokamak field
$B_p$ of several Tesla or more, it is important to explore the regime
$S_r < 1$, which is the primary focus of  the remainder of this paper.
We have identified this regime ({\emph{right} panel of Figure~\ref{nonlocal}) as a promising one for CT
fueling of ITER-relevant plasmas due to the precise spatial deposition
of  the CT and the deep penetration that can be achieved for core
fueling applications. In this paper, we have focused on background plasmas with $\beta_p=0.02$ like in {\it ITER}, while simulation results with $\beta_p=0.2$ like in the National Spherical Torus Experiment (NSTX) will be studied in follow-on work.

\subsection{CT Evolution}\label{overview}

In this sub-section, we describe in detail the CT evolution for the
ITER-relevant regime of $V_{\rm L}<v_{\rm inj}\lesssim V_{\rm AP}$
and $S_r < 1$.  In this regime, the CT evolution can be divided into
three stages: (1)~initial CT penetration, (2)~CT compression in the
propagation direction ($z$) and reconnection, and (3)~coming to rest
and spreading in the $x$ (toroidal) direction.  Magnetic reconnection,
starting late in the second stage in our simulations, arises due to
numerical diffusion and leads to mixing between the CT and background
plasmas.

In order to mimic CT injection with ITER-relevance, we adopt physical
quantities as given in Table~\ref{iter} and choose the injection speed
$v_{\rm inj}$ such that $V_{\rm AC}=1.0<v_{\rm inj}=1.1<V_{\rm
AP}=3.16$ (this falls into the ``ITER-relevant'' case of
Figure~\ref{scheme}). Figure~\ref{evolution} displays the time evolution
of the plasma density (color contours in common logarithmic scale) in
the $x$-$z$ plane with $y=0$. The white solid contour lines indicate
the magnetic pressure $p_B=B^2/2$. Fig.~\ref{axial_disp} displays the axial profile of $x$ and $y$ integrated density $\int_x \int_y \rho dxdy$ at different times corresponding to Fig.~\ref{evolution}. The entire evolution can be
described by the three stages mentioned above.

Figure~\ref{reconnection} displays the magnetic field $B_{xz}$,
(arrows) and current density $j_y$ (color contours) in the $x$-$z$ plane at
$y=0$. During the first stage (initial penetration), the CT
experiences a very strong magnetic obstacle, and therefore the CT
plasma is highly compressed and the plasma density increases at the
interface between the CT and background plasmas. A large plasma
current also appears at the interface due to the compression of the
background magnetic field seen in Figure~\ref{reconnection}(a). This
current sheet is bent and broken into two parts
(Figure~\ref{reconnection}(b)) due to the magnetic field configuration
of the CT field. Some reconnection takes place at the left part of the
current sheet as shown in Figure~\ref{mechanism}. The CT is successively
decelerated by the magnetic tension force of the background magnetic
field.

After the CT has fully entered the background plasma region (after
$t\sim4$), the background field lines, some of which have reconnected
with the CT magnetic field at the CT's leading edge, are reconnected
again with the CT magnetic field at the CT's trailing edge, as seen in
Figure~\ref{reconnection}(c). As pointed out by Suzuki et al.\cite{shk00}, via this
process the CT is separated from the background fields lines, through
which the magnetic tension force decelerating the CT is relaxed. At
$t=4.375$, a magnetic configuration schematically shown in
Figure~\ref{mechanism} is formed.  This interpretation is supported by
Figure~\ref{mech}, which displays the axial distributions of density and
$B_x$.  From Figure~\ref{mech}, we can see that $B_x$ changes sign from
positive to negative and then to positive again. The transition from
positive to negative happens at larger $z$ on
the left hand side while at smaller $z$ on the right hand side, which is the case in Figure~\ref{mechanism}(\emph{right})
(see two dashed lines in Figure~\ref{mechanism}(\emph{right})). From
Figure~\ref{mechanism}, primary reconnection sites are at the upper left
and lower right sections of the CT.  The reconnection process allows
the high-density CT plasma to escape from the CT and eventually flow
outward along the background magnetic field horizontally
(Figure~\ref{mechanism}). The CT plasma starts to contract in $z$ and
expand in $x$. The reconnection is asymmetric about the CT axis, and
this asymmetry results in CT plasma outflow in a direction that is not
completely in the $x$ direction, but rather obliquely (panel (d) of
Figure~\ref{reconnection}). Eventually it will become more parallel to
$x$ (\emph{right} panel of Figure~\ref{nonlocal}). Magnetic flux is being destroyed as well. During this process the
MHD wave drag from Alfv\'en waves induced by horizontal plasma outflow
\cite{pp88,nw91} might further slow down the CT plasma. Note that
Suzuki et al.\cite{shk01} showed that the CT penetration depth, based on a model
with magnetic tension force as the main deceleration mechanism,
matches simulation results very well, implying that MHD wave drag
forces may not be important in CT deceleration.  The initially injected
magnetic and perpendicular kinematic energies are converted into
parallel kinematic energy. Contrary to \cite{pp88} and
\cite{shk00,swsh00,shk01}, CT tilting, the time scale of which is
proportional to $\sqrt{B_bB_p/\rho_b}$, is not observed in our case
due to the fast injection and short CT transit time as required for
the ITER-relevant regime (Table~\ref{iter}).

After the high-density CT plasma has been depleted during the
compression stage (after $t\sim9$), the CT field diminishes and only
the slightly perturbed background field survives, almost orienting to
the direction of the initial background magnetic field (see
Figure~\ref{evolution} and Figure~\ref{reconnection}(d)). Almost all of the
initially injected magnetic and kinematic energy now survive in the
form of parallel kinematic energy. The CT comes to rest in the
injection direction ($z$). A narrow elongated structure along $x$
results, as seen in the lower middle and lower right panels of
Figure~\ref{evolution}. This line-shaped structure with a spread of only
$\Delta z \sim0.2$  (much smaller than the size of the background
plasma, see also Fig.~\ref{axial_disp})  implies a very precise fuel deposition capability.

\subsection{CT penetration depth}\label{depth}

In this sub-section, we establish the dependence of CT penetration
depth, an important parameter for CT fueling of tokamaks, on the
experimentally controllable  parameters of CT injection speed $v_{\rm
inj}$  and initial density $\rho_b$, which collectively determine the initial
CT energy.  The penetration depth $S$ is defined as the axial ($z$) distance
between the final mean position of the injected CT and the injection
location on the boundary.  We find that $S$ is highly dependent on the
initial injection speed $v_{\rm inj}$ and CT density
$\rho_b$. Figure~\ref{penetration}(a) displays the relationship between
the $S$ and $v_{\rm inj}$, showing that  $S$ is proportional to
$v_{\rm inj}$ if $V_{\rm L}<v_{\rm inj}< V_{\rm AP}$.
Figure~\ref{penetration}(b) displays the relationship between the CT
density and penetration depth, which shows that the penetration depth
$S$ increases with the CT density. The CT spread in the $z$ direction
is around $\sim0.2$ in all cases (see also Fig.~\ref{axial_disp}), which is very small compared to the
background plasma size $\sim20$. These two empirical relationships are
important in the sense that they provide clues for how to choose the
injection speed and CT density to get precise fuel deposition,
therefore controlling the core plasma profile in a large tokamak such
as {\it ITER}. It is very hard to manage that by other methods such as
pellet injection \cite{om08}.

For the case shown in Figure~\ref{penetration}(b), the parameters given
in Table~\ref{iter} are adopted ($v_{\rm inj}=2.0$) except CT density $\rho_b$ and CT
field $B_b$. We choose them such that the plasma  parameter $\beta=0.2$ is
kept constant for all the simulation results shown in
Figure~\ref{penetration}(b).   We keep $\beta$ constant since
experimentally this is more reasonable than changing CT density or CT
field strength independently. Thus, increasing CT density means
increasing both the initial kinematic energy $E_k=\int1/2\rho_b v_{\rm
inj}^2 dV$, where $dV$ is the infinitesimal CT volume, and initial
magnetic energy $E_m=\int 1/2B_b^2dV$. However, because of the
limiting criterion given in Eq.~\ref{cs}, the initial CT kinematic energy $E_k$
should be much larger than the initial CT magnetic energy $E_m$ since
the background field is much larger than the CT field in all
simulation results presented in Figure~\ref{penetration}(b) ($B_b/B_p=0.1$). Also the residual initial
total force $\mathbf{j}\times\mathbf{B}-\nabla p$ is proportional to
the CT density $\sim O(\rho_b)$. Therefore, increasing the CT density
$\rho_b$ would elevate the initial residual force, which would result
in shock/wavefronts. Thus, with too large a density ratio
($\rho_b/\rho_p\gtrsim15$, which is inferred from simulation results), the fueling is not as localized as the
case with smaller CT density, while generating similar phenomena as
the shock/wavefront dominated cases.

Results of Figure~\ref{penetration} remind us of the importance of
initially injected energy upon the penetration depth.
As discussed in \S\ref{overview}, how long the compression stage lasts
is determined by the sum of the initially injected kinematic and
magnetic energy. The larger the sum is, the longer it will take for
the reconnection to dissipate/convert the energy. If we assume that
the deceleration mechanism of the CT is mainly due to the magnetic
tension force, which is independent of the initially injected
energy \cite{shk01}, then the larger initially injected energy would
lead to deeper fueling given the same injection speed $v_{\rm inj}$
(since plasma parameter $\beta\ll1$, the internal energy of the CT
plasma can be ignored). Certainly the larger injection speed would
lead to deeper penetration.

Figure~\ref{decay} shows the time evolution of the net toroidal magnetic
flux $\psi_t=\int B_y dS$ (only positive $B_y$ is selected) with
$v_{\rm inj}=1.1$ and $\rho_b/\rho_p=10$. Before $t=6.25$, the net CT
toroidal magnetic flux increases due to the initial penetration and
compression of the magnetic field lines at the interface of the CT and
background plasma. After $t=11.25$, the line structure reaches the
horizontal boundaries and some toroidal magnetic flux flows
out. Between $t=7.5$ and $t=11.25$, the only source of the destruction
of the net toroidal magnetic flux is magnetic reconnection due to the
numerical diffusion. If we fit the decay as
$\psi_t(t)/\psi_t(t=0)\equiv\exp(-t/\tau_{\rm res})$, the ``resistive"
dissipation time due to numerical diffusion is $\tau_{\rm res}\sim
10.9$. Therefore the global resistive decay due to numerical diffusion is not important on the time
scales of CT transit time $(t\sim5)$ given injection speed $v_{\rm
inj}=1.1$. In the simulation, mixing arises from numerical diffusion induced reconnection and occurs on the same time scale as the CT transit time. However, the mixing should be less of a factor in reality due to the much smaller diffusion in a high temperature tokamak.


\section{Conclusions \& discussion}\label{conclusion}
In this paper we presented nonlinear ideal MHD simulation
results of a compact toroid injected into a hot strongly magnetized
plasma. The simulations are intended to provide
insights into CT fueling of an ITER-class tokamak.
As a first step, we have investigated the
problem of a high density low $\beta$ CT injected into a slab
background plasma with uniform magnetic field and density.  We intend to
investigate the injection of
high $\beta$ CT's as well as unmagnetized dense plasma jets,
and also incorporate more realistic background profiles
in follow-on work.

Our main findings are as follows.  A regime is identified in terms of
CT injection speed and CT-to-background magnetic field ratio that
appears promising for precise core fueling.
Shock-dominated regimes, which are probably unfavorable for
tokamak fueling, are also identified.  The CT penetration depth is proportional to the CT
injection speed and density.  For the regime identified as
favorable for precise core fueling, the entire CT evolution can be
divided into three stages:  (1)~initial penetration, (2)~compression in the
direction of propagation, and reconnection with the background plasma, and (3)~coming to rest and spreading in the
direction perpendicular to injection.  Tilting of the CT is not
observed due to the fast transit time of the CT across the background plasma.
Reconnection occurring at the upper left and lower right
portions of the CT fragments the CT and leads to CT plasma outflow
horizontally along the background magnetic field lines, forming a line-shaped
structure almost parallel to the background magnetic field. We have also
studied how the penetration depth is determined and have found an empirical
relationship between the penetration depth versus the initial injection speed
and CT plasma density. The penetration depth is proportional
to the injection speed and CT density.  When the CT comes to rest, it is
highly compressed in the direction of injection, implying
that very precise deposition is possible.

In a device with high bootstrap current fraction, optimized density
and pressure profiles must be maintained. A core fueling system is all that a
burning plasma may have to alter core plasma
conditions and for burn control. Initial density peaking via core
fueling provides more flexibility to reach ignition.
If a CT injection system
is within the ITER-relevant, no-shock/wavefront regime as required by
$v_{\rm inj}\ll V_{\rm AP}$, $\rho_b/\rho_p\lesssim15$ and
$B_b/B_p\ll1$, then precise core deposition with
minimal background equilibrium perturbation may be possible.

Because $V_{\rm AC} \sim1700\;\unit{km\;s^{-1}}$ for ITER (using parameters in Table~\ref{iter}), it will be a challenge for present CT injectors with $v_{\rm inj} \sim 400\;\unit{km\;s^{-1}}$ to achieve deep penetration.  A potential solution is to increase $\rho_b/\rho_p$ to at least $100$, which would reduce $V_{\rm AC}$ to a more reasonable value ($\sim170\;\unit{km\;s^{-1}}$).  Due to computational limitations, a simulation of such a case (very dense CT) is beyond the scope of this study.  However, to provide insight into the penetration characteristics of a very high density injected plasma, we report here some preliminary results on the injection of a dense ($\rho_b/\rho_p=538$ or $\rho_b\sim10^{17}\;\unit{cm^{-3}}$), cool ($T_b/T_p=3.33\times10^{-4}$ or $T_b=2.5\;\unit{ev}$), and very weakly magnetized plasma \cite{wfd09}, which would be an interesting alternative to CT injection.  Figure~\ref{high} suggests that this type of injected plasma is capable of reaching the core of the tokamak, albeit with worse localization as discussed in Sec.~\ref{depth}.  This will be reported on in more detail in follow-on work.

\ack
The authors thank Shengtai Li for extensive advice on the code.  The authors also thank Roger Raman and Xianzhu Tang for very useful discussions and constructive comments. 
This work was funded by DOE contract no.DE-AC52-06NA25396 under the Los Alamos Laboratory
Directed Research and Development (LDRD) Program.



%
\begin{figure}[!htp]
\begin{center}

  \scalebox{0.4}{\includegraphics{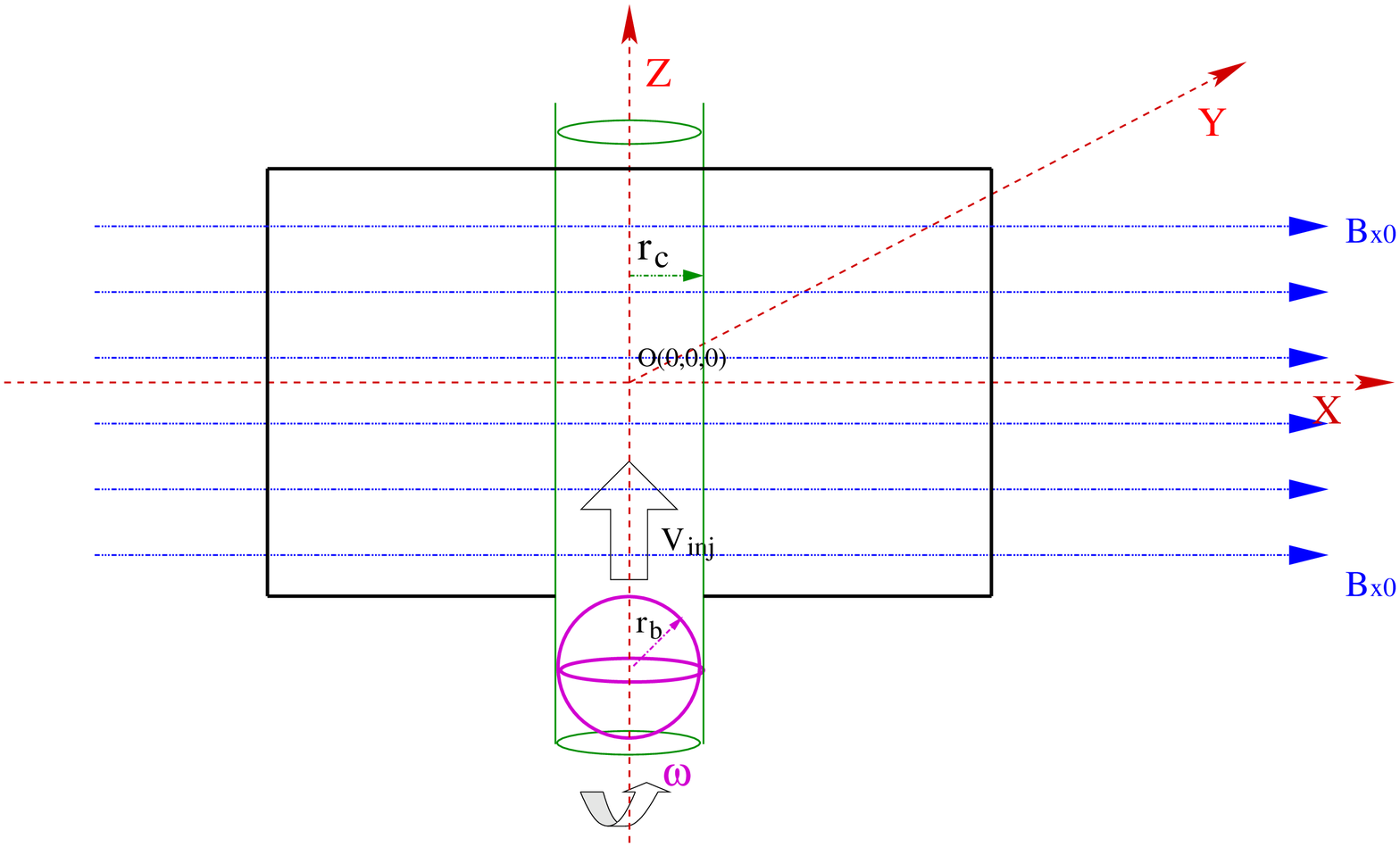}}
  \caption{\label{config}~Schematic of the simulation geometry showing the coordinate system. In the text, the direction along the $z-$axis is defined as the axial direction.}
\end{center}
\end{figure}

\begin{figure}[!htp]
\begin{center}

  \scalebox{0.4}{\includegraphics{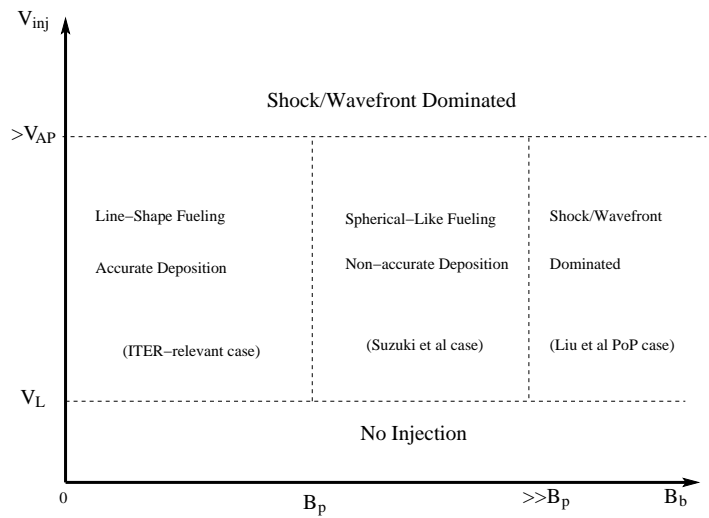}}
  \caption{\label{scheme} Qualitative behavior for CT injection in terms of CT injection speed and magnetic field strength.}
\end{center}
\end{figure}

\begin{figure}[!htp]
\begin{center}
  \scalebox{0.12}{\includegraphics{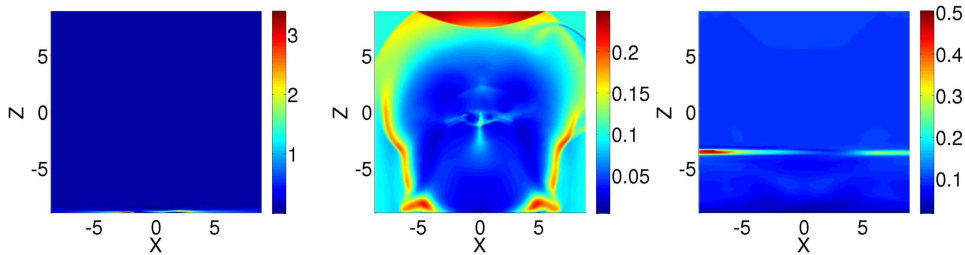}}
  \caption{\label{nonlocal}~(color) Density in the $x$-$z$ plane at $y=0$. \emph{Left} panel: at $t=20$ with injection speed $v_{\rm inj}=0.3$, which shows no penetration.   \emph{Middle} panel: at $t=2.5$ with injection speed $v_{\rm inj}=10.0$, which shows non-localized deposition. \emph{Right} panel: at $t=20$ with injection speed $v_{\rm inj}=1.1$, which shows highly localized deposition. $V_{\rm AP}=3.16$.}
\end{center}
\end{figure}

\begin{figure}[!htp]
\begin{center}

  \scalebox{0.8}{\includegraphics{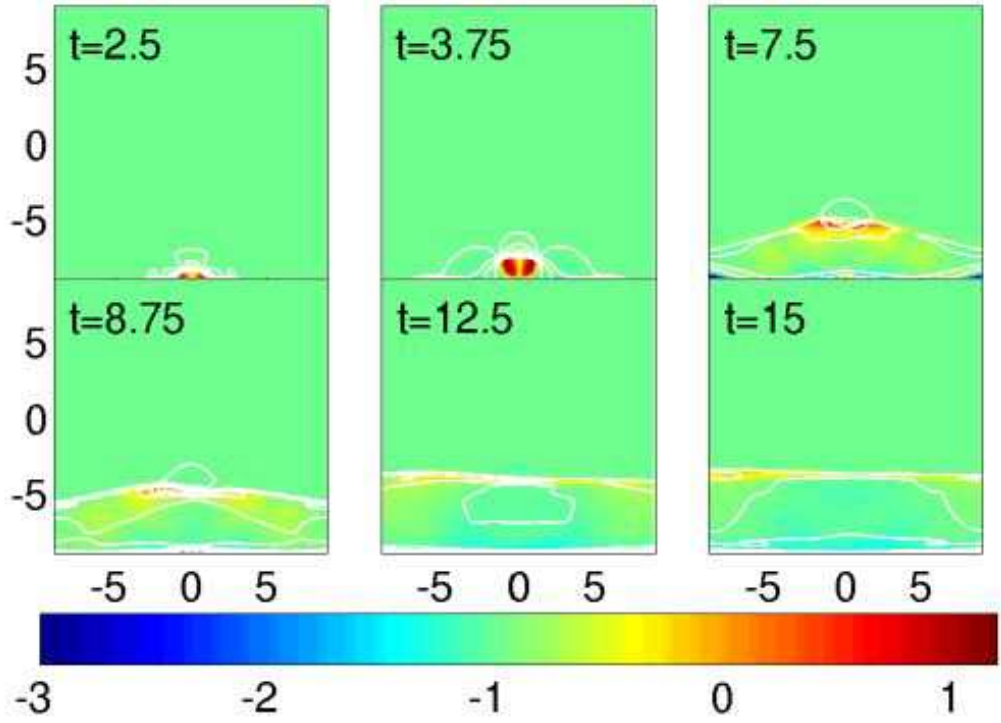}}
  \caption{\label{evolution}~(color) Density (logarithmic scale) in the $x$-$z$ plane at $y=0$ as a function of time. The white lines are contours of  magnetic pressure $p_B=B^2/2$. For $t=2.5$, $p_B\in[0.0,1.6843]$; for $t=3.75$, $p_B\in[0.0,1.7612]$; for $t=7.5$, $p_B\in[0.0,1.4405]$; for $t=8.75$, $p_B\in[0.0,1.1872]$; for $t=12.5$, $p_B\in[0.0,1.0219]$; for $t=15$, $p_B\in[0.0,1.008]$.}
\end{center}
\end{figure}

\begin{figure}[!htp]
\begin{center}
  \scalebox{0.4}{\includegraphics{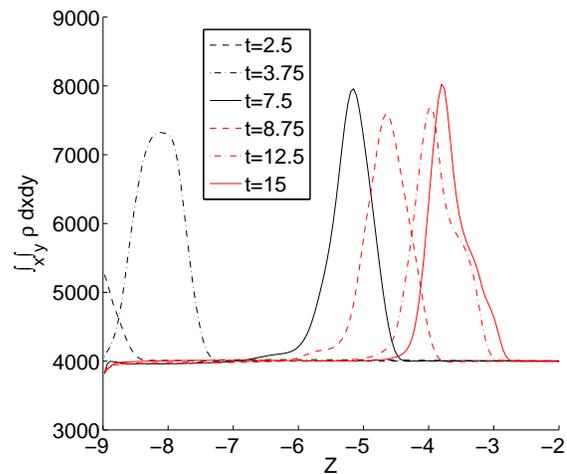}}\;  
  \caption{\label{axial_disp}~(color) $x$ and $y$ integrated density $\int_x \int_y \rho dx dy$ versus $z$ at different times corresponding to Fig.~\ref{evolution}. Black dash: $t=2.5$, black dash dot: $t=3.75$; black solid: $t=7.5$; red dash: $t=8.75$; red dash dot: $t=12.5$, red solid: $t=15$.}
\end{center}
\end{figure}

\begin{figure}[!htp]
\begin{center}
  \scalebox{0.2}{\includegraphics{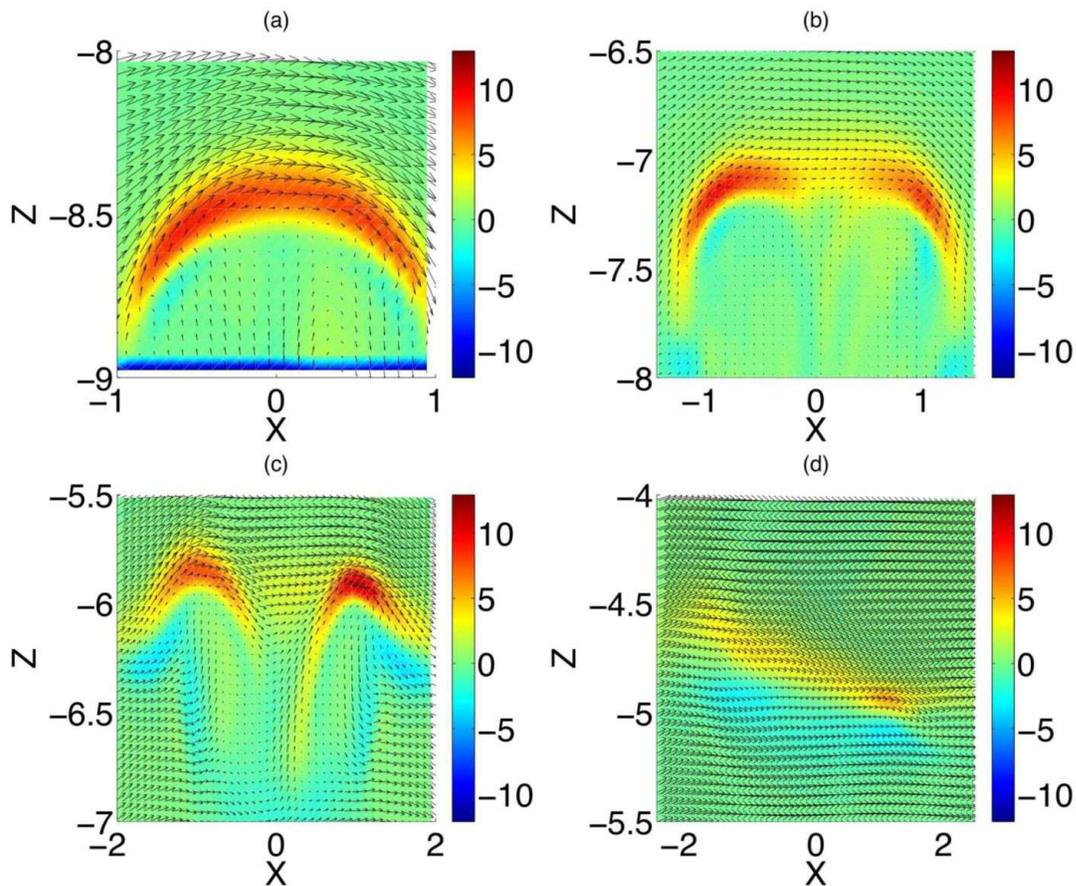}}\;  
  \caption{\label{reconnection}~(color) Magnetic field and current density $j_y$ in the $x$-$z$ plane as a function of time at $y=0$. The color contours show $j_y$ while arrows show $B_x$ and $B_z$, which are normalized to their maximum values. (a) $t=2.5$; (b) $t=4.375$; (c) $t=6.25$; (d) $t=8.75$. Please note that the scales for the abscissa and ordinate are not identical.}
\end{center}
\end{figure}

\begin{figure}[!htp]
\begin{center}

  \scalebox{0.15}{\includegraphics{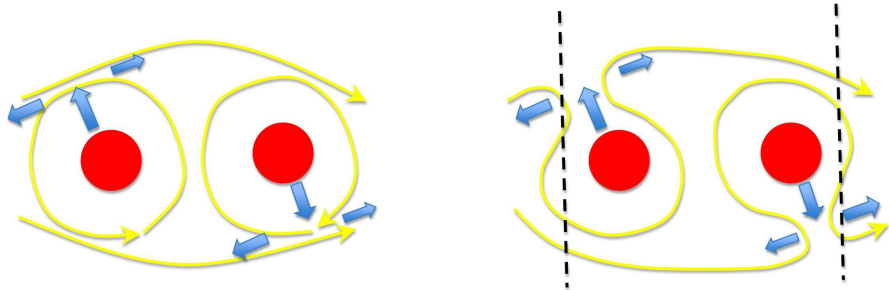}}
   \caption{\label{mechanism}~(color) Diagram of the magnetic configuration in the contraction stage.  Red color indicates the high-density CT plasma. Yellow lines indicate magnetic field lines, while blue arrows indicate CT plasma flow patterns. See Figure~\ref{mech} for dotted black lines.}
\end{center}
\end{figure}

\begin{figure}[!htp]
\begin{center}

  \scalebox{0.15}{\includegraphics{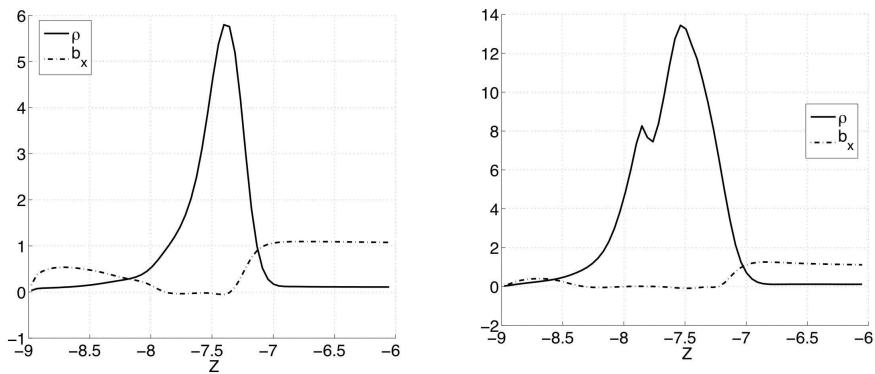}}
   \caption{\label{mech}~Axial profiles of several quantities at $t=4.375$ with $(x,y)=(-1.125,0)$ (\emph{left}) and $(x,y)=(0.855,0)$ (\emph{right}) . The density $\rho$ and magnetic field strength in the $x$-direction $B_x$ are shown for evaluating the magnetic configuration shown in Figure~\ref{mechanism}(\emph{right}). These two axial profiles correspond to the two dash lines in Figure~\ref{mechanism}(\emph{right}).}
\end{center}
\end{figure}

\begin{figure}[!htp]
\begin{center}

  \scalebox{0.4}{\includegraphics{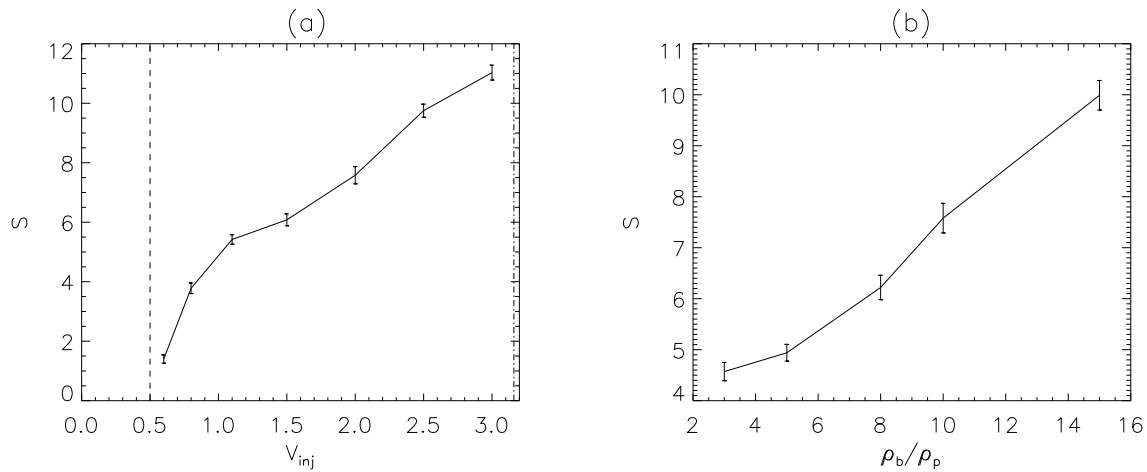}}
  \caption{\label{penetration}~(a) Penetration depth $S$ versus injection speed $v_{\rm inj}$ with $\rho_b/\rho_p=10$ and (b) Penetration depth $S$ versus CT density $\rho_b$ with $v_{\rm inj}=2.0$. Both panels assume $B_b/B_p=0.1$. The penetration depth is defined as the distance between the final mean position of the injected plasma and the bottom boundary.  The error bar indicates the final size of the injected CT in the $z$-direction. In panel (a), the dash line indicates $V_{\rm L}\sim0.5$ and the dash dot line indicates $V_{\rm AP}=3.16$.}
\end{center}
\end{figure}

\begin{figure}[!htp]
\begin{center}

  \scalebox{0.4}{\includegraphics{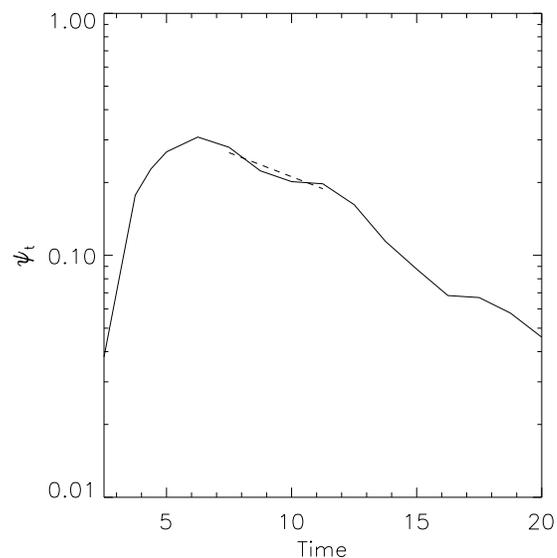}}
  \caption{\label{decay}~Decay of the net toroidal magnetic flux $\psi_t=\int B_y dS$, where only positive $B_y$ is selected with $v_{\rm inj}=1.1$ and $\rho_b/\rho_p=10$. The dashed line fits the data between $t=7.5$ and $t=11.25$ with the formula $\psi_t(t)/\psi_t(t=0)\equiv\exp\left(-t/\tau_{\rm res}\right)$, where $\tau_{\rm res}=10.9$ is the resistive dissipation time due to numerical diffusion, and it is much larger than the CT transit time $\sim5$ given injection speed $v_{\rm inj}=1.1$.}
\end{center}
\end{figure}

\begin{figure}[!htp]
\begin{center}
  \scalebox{0.5}{\includegraphics{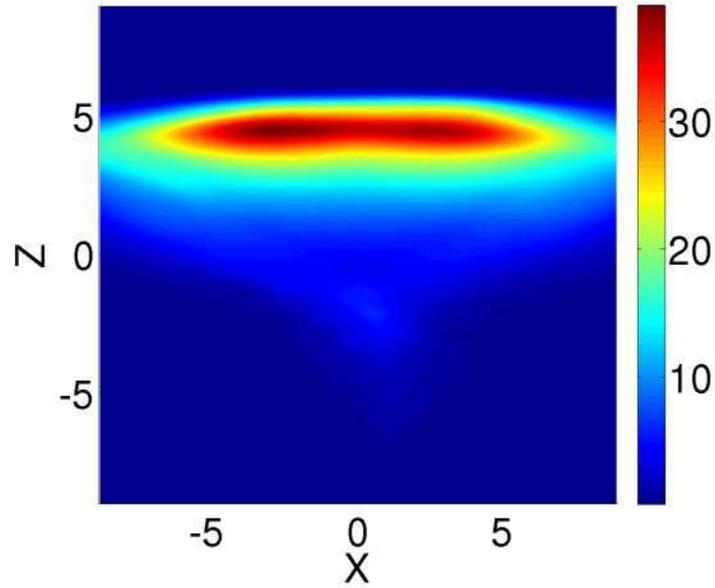}}
  \caption{\label{high}~(color) Density in the $x$-$z$ plane of $y=0$ at $t=150$ with injection speed $v_{\rm inj}=0.12$, $\rho_b/\rho_p=538$, and $B_b/B_p=0.001$, which shows worse localized deposition. Please note that the CT is still moving with almost the initial injection speed.}
\end{center}
\end{figure}

%


%
\begin{table}[!htp]
\caption{\label{iter}~Normalized physical quantities. Note that $V_{\rm AC}=B_{p}/\sqrt{\rho_b}=1.0$. Injection speed $v_{\rm inj}$ is between $V_{\rm L}$ and $V_{\rm AP}$. Plasma is assumed to be composed of half deuterium and half tritium. The plasma density is the sum of electron and ion densities. }
\begin{tabular}{ccccc}

\hline 
\hline
&
 \multicolumn{2}{ c }{ CT} &
  \multicolumn{2}{c }{ Background} \tabularnewline
  \hline
  Physical Quantities &
  numerical &
  physical &
  numerical &
  physical \tabularnewline
\hline
Magnetic Field &
$B_b=0.1$ &
$0.53\;\unit{T}$ &
$B_p=1.0$ & 
$5.3\;\unit{T}$ \tabularnewline
Density &
$\rho_b=1.0$ &
$1.86\times10^{15}\;\unit{cm^{-3}}$&
$\rho_p=0.1$ &
$1.86\times10^{14}\;\unit{cm^{-3}}$ \tabularnewline
Temperature &
$T_b=0.001$ &
$75\;\unit{ev}$ &
$T_p=0.1$ &
$7.5\;\unit{kev}$ \tabularnewline
plasma $\beta=2\rho T/<B^2>$&
\multicolumn{2}{ c }{ $\beta_b=0.2$} &
\multicolumn{2}{ c }{ $\beta_p=0.02$} 
\tabularnewline
Alfv\'en Speed $V_{\rm A}=B/\sqrt{\rho}$&
$V_{\rm AB}=0.1$&
$170\;\unit{km\;s^{-1}}$ &
$V_{\rm AP}=3.16$&
$5.4\times10^3\;\unit{km\;s^{-1}}$\tabularnewline
Sound Speed $V_c=\sqrt{\gamma T}$&
$V_{\rm cB}=1.7\times10^{-3}$&
$2.89\;\unit{km\;s^{-1}}$ &
$V_{\rm cP}=0.41$ &
$7.0\times10^{2}\;\unit{km\;s^{-1}}$\tabularnewline
\hline
\hline
\end{tabular}
\end{table}

\end{document}